\documentclass[a4paper, 10pt, conference]{ieeeconf}      % Use this line for a4 paper

\IEEEoverridecommandlockouts                              % This command is only needed if
\usepackage{amsmath} % assumes amsmath package installed
\usepackage{amssymb}  % assumes amsmath package installed

\usepackage{cite}
\usepackage{graphicx}
\usepackage[caption=false, font=footnotesize]{subfig}
\usepackage{placeins}
\usepackage{tikz}
\usepackage{pgfplots}

\usepackage{algorithm}
\usepackage{algorithmic}

\newtheorem{theorem}{Theorem}

\newcommand\copyrighttext{%
	\footnotesize Copyright $\copyright$ 2019 IEEE.
	Personal use of this material is permitted.
	Permission from IEEE must be obtained for all other uses, in any current or future media, including reprinting/republishing this material for advertising or promotional purposes, creating new collective works, for resale or redistribution to servers or lists, or reuse of any copyrighted component of this work in other works.}%
\newcommand\copyrightnotice{%
	\begin{tikzpicture}[remember picture,overlay]%
	\node[anchor=south,yshift=10pt] at (current page.south) {\fbox{\parbox{\dimexpr\textwidth-2cm}{\copyrighttext}}};%
	\end{tikzpicture}%
	\vspace{-10pt}%
}

\title{\LARGE \bf
	A Risk and Comfort Optimizing Motion Planning Scheme for Merging Scenarios$^\ast$
}

\author{Johannes M{\"u}ller$^{1}$ and Michael Buchholz$^{1}$% <-this % stops a space
	\thanks{*This work was financially supported by the Federal Ministry of Economic Affairs and Energy of Germany within the program "Highly and Fully Automated Driving in Demanding Driving Situations" (project MEC-View, grant number 19A16010I).}% <-this % stops a space
	\thanks{$^{1}$Johannes M{\"u}ller and Michael Buchholz are with the Institute of Measurement, Control and Microtechnology,
		Ulm University, D-89081 Ulm, Germany
		{\tt\small \{johannes-christian.mueller, michael.buchholz\}@uni-ulm.de}
	}%
}

\begin{document}

	\maketitle
	\copyrightnotice
	\thispagestyle{empty}
	\pagestyle{empty}

	%%%%%%%%%%%%%%%%%%%%%%%%%%%%%%%%%%%%%%%%%%%%%%%%%%%%%%%%%%%%%%%%%%%%%%%%%%%%%%%%
	\begin{abstract}
		
	Motion planning for merging scenarios accounting for measurement and prediction uncertainties is a major challenge on the way to autonomous driving. Classical methods subdivide the motion planning into behavior and trajectory planning, thus narrowing down the solution set. Hence, in complex merging scenarios, no suitable solution might be found. In this work, we present a planning scheme that solves behavior and trajectory planning together by exploring all possible decision options. A safety strategy is implemented and the risk of violating a safety constraint is minimized as well as the jerk to feature a risk and comfort optimal trajectory. To mitigate the injection of noise into the actual trajectory, a new analytical trajectory generation method is derived and its optimality is proven. The decision capability is evaluated through Monte-Carlo simulation. Furthermore, the calculation time is evaluated showing the real-time capability of our approach.
		
	\end{abstract}

	%%%%%%%%%%%%%%%%%%%%%%%%%%%%%%%%%%%%%%%%%%%%%%%%%%%%%%%%%%%%%%%%%%%%%%%%%%%%%%%%
	\section{INTRODUCTION}
	Motion planning for merging scenarios is a major challenge on the way to autonomous driving. Accident statistics \cite{Simon2009} suggest that even among human drivers, intersection scenarios are challenging: other vehicles have to be predicted, which inherently comes with uncertainties that need to be accounted for, the possible decisions need to be explored, and a decision has to be made for the most suitable option including an adequate timing. For a technical system, all of this has to be performed with high reliability under strict real-time constraints.
	
	Classical approaches subdevide the problem into behavior planning and trajectory planning and solve the problems individually \cite{Paden2016}. However, due to the missing feedback, the possible trajectories have to be restricted, e.g. by preplanned trajectories \cite{Puphal2018} to maintain feasibility for the consequent trajectory planner \cite{Hubmann2018a}. In turn, Partially Obeservable Markov Decision Process (POMDP)-based or similar methods use reinforcement learning \cite{Gritschneder2016} or related methods \cite{Hubmann2018a} to holistically solve the motion planning problem by learning an appropriate control law.
	
	We address the problem with a sampling-based planning scheme that explores the available merging options in terms of target states including the respective arrival time. Uncertainties are modeled in a probabilistic way and the residual risk of violating a safety constraint is minimized as it is accounted for in the overall cost function. The ego state is connected with the target states using a new analytical time-weighted jerk optimal trajectory generation method that mitigates the injection of noise into the actual trajectory. For the trajectory generation method, optimality with respect to the problem formulation is proven. During the exploration, a constraint check is directly applied to all solution candidates guaranteeing feasibility. Furthermore, a fail-safe strategy is implemented for all possible solutions, thus, in case no feasible solution is available at all, the vehicle can still be transfered into a safe state at reduced passengers' comfort. From all valid options, the globally optimal solution is chosen. Thus, the motion planning scheme holistically decides for the ego vehicle's tactical behavior, the timing, and the corresponding trajectory.
	
	The contributions of this paper are twofold: First, we present a new planning scheme that holistically considers multiple options, accounts for uncertainties by calculating the residual risk of violating a safety constraint, and optimizes for this risk as well as for the passengers' comfort by minimizing risk and jerk. Secondly, we derive a new analytical trajectory generation method that creates time-weighted jerk optimal trajectories. We show that the resulting trajectories are optimal with respect to the problem formulation and demonstrate its benefits through simulation. We further evaluate the merging decisions of the planning scheme through Monte-Carlo simulation showing its ability to find reasonable decisions. Finally, we evaluate the calculation time showing good real-time capability of the method.
	
	\subsection{Related Work}
	The related work can be roughly classified in four categories: the classical approaches, the set-based methods, the approaches based on Markov decision processes, and the communication-based approaches.
	
	The latter show very promising results, in particular, as the other vehicle's intent and future trajectory do not need to be estimated, but are communicated to or even negotiated with the ego vehicle. However, it is assumed that the vehicles are interconnected, which is not the case for current mixed traffic. We refer to the surveys \cite{RiosTorres2017} and \cite{Pereira2017} for details about this category.
	
	Classical approaches that subdivide the motion planning into behavior and trajectory planning \cite{Paden2016} have the advantage of good computational tractability and modularity \cite{Hubmann2018a}. However, through the neglected feedback, the set of solution is reduced. Thus, in difficult planning situations, no solution can be found \cite{Hubmann2018a}. Recent examples for the classical approach which account for uncertainties are \cite{Tas2018} and \cite{Puphal2018}.
	
	Set-based methods, e.g. \cite{Pek2017} or \cite{Pek2018}, are mainly developed by the research group around Althoff and focus on the safety aspect trying to give guarantees that the intended maneuver is safe. The drawback of these methods is that they tend to be overly conservative. Thus, in dense traffic and a highly uncertain environment, these methods are likely to yield no suitable solution.
	
	Finally, approaches based on Markov Decision Processes use learning data and mostly reinforcement learning \cite{Gritschneder2016} or related methods \cite{Hubmann2018a} to learn an appropriate control law guiding the ego vehicle through the merging scenario. The approaches of this category holistically solve the problem, inherently accounting for uncertainties. Given the appropriate learning data, the same method can be used for different traffic scenarios. However, lots of learning data is necessary to learn the appropriate control law. Furthermore, these methods suffer from a high computational complexity, thus approximate solutions have to be found. Thus, due to an abstraction of goals and states, these works perform "utmost worse"\cite{Krishnan2018} compared to classical approaches. Further works from this field are, e.g., \cite{Sezer2015} or \cite{Evestedt2016}.
	
	In contrast to \cite{Hubmann2018a}, we consider the interaction between other road users and the ego vehicle as a constraint,
	i.e. the trajectory is optimized, besides for the jerk, only for the risk of violating these constraints.
	Thus, we argue that for our problem formulation, the prediction of the other road users is only loosely coupled with the planning and separating prediction and planning thus results in an insignificant reduction of planning accuracy. However, due to the decoupling, the computational complexity is reduced.
	
	\section{PLANNING SCHEME}
	In this section, the planning scheme is presented. Starting from a brief description of our automated vehicle, the problem formulation is given, followed by an overview on the algorithm. Finally, the algorithm steps are explained in detail.	

	Our goal is to find a safe merging trajectory $\mathcal{T}$ that takes measurement and prediction uncertainty into account, minimizes the residual risk of getting into a dangerous situation, and maximizes the comfort of the passengers.
	As we consider merging scenarios on streets, i.e. in highly structured environments, we reduce the planning problem to 1D by \textit{path velocity decomposition}, which is a common strategy \cite{Paden2016}, \cite{Hubmann2018a}. By using \textit{path velocity decomposition}, the planning is abstracted from the geometry of the merging scenario. Thus, our approach can be applied to various merging scenarios such as ramp merging in a high way scenario, merging at a narrowing, or merging into an unsignalized urban yield intersection. In this work, however, for the sake of clear and intuitive explanation, we focus on the scenario of merging into an unsignalized urban yield intersection. We furthermore assume that no other vehicle is between the ego vehicle and the merging point blocking the ego vehicle from merging freely into a traffic gap on the main road.
	
	\subsection{Automated Vehicle Architecture}
	The architecture of our automated vehicle, i.e. the ego vehicle, basically consists of four layers: the sensors measure the vehicle's environment, the perception layer processes this information and summarizes it in the environment model. This, in turn, is handed over to the motion planning module. The motion planning module is subdevided into a a prediction module that features predictions to the current objects $\mathcal{P}_o$ in the object list $\mathcal{L}_o$, and planning scheme, which is addressed in this work.
	The trajectory generated by the motion planning module finally is passed to a subordinate control layer that stabilizes the vehicle on the road and controls the correct execution of the motion plan through the actuators. The environmental model can optionally be enriched by extern cooperative information that is communicated to the ego vehicle through Vehicle-to-Anything (V2X) communication.
	
	\subsection{Problem Formulation} \label{sec:ProblemFormulation}
\begin{figure}
		\centering
		\includegraphics[width=0.3\textwidth]{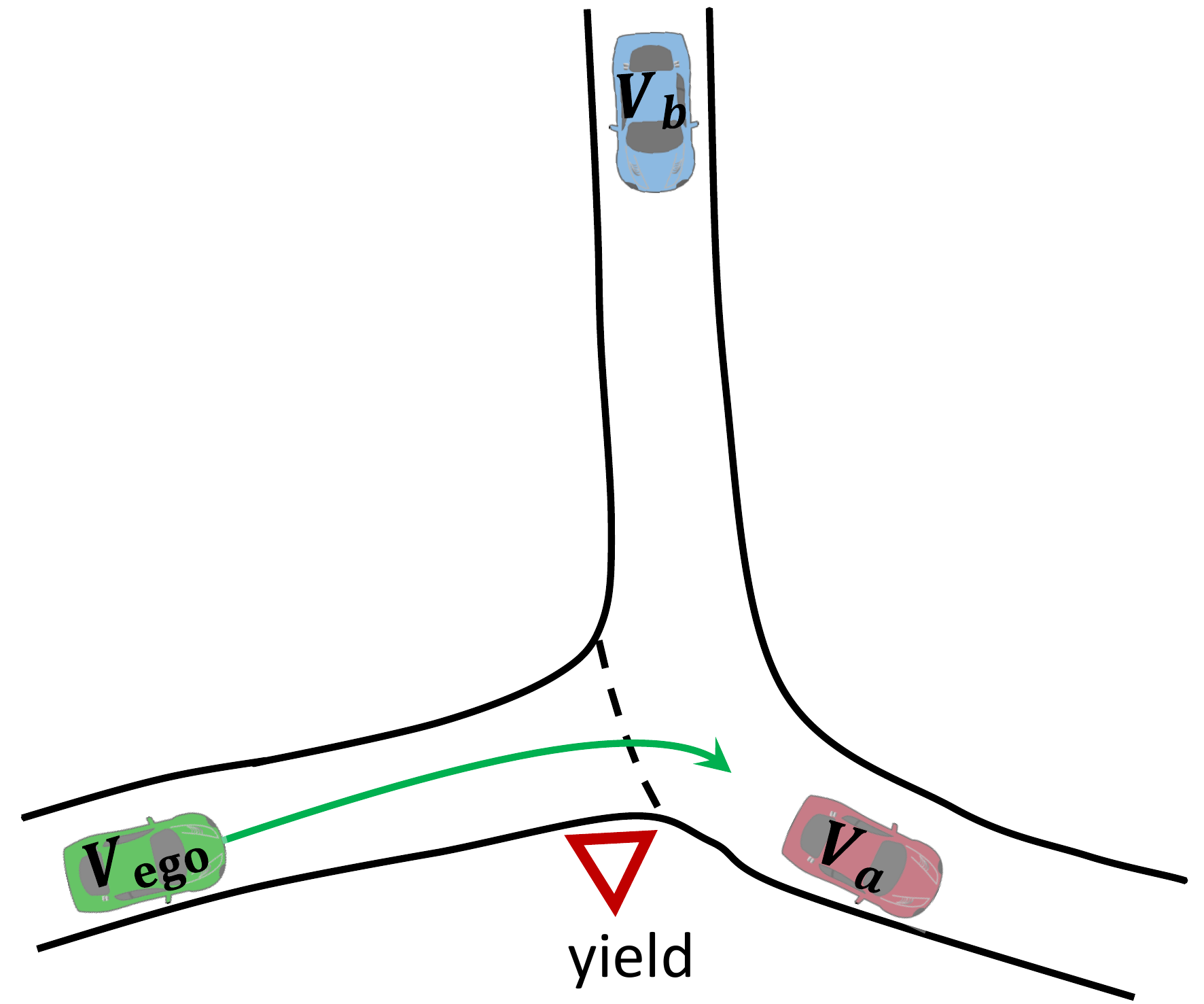}
		\caption{Traffic scenario. The ego vehicle $\text{V}_\text{ego}$ tries to merge while two other vehicles $\text{V}_a$ and $\text{V}_b$ approach the intersection.}
		\label{fig:Scenario}
	\end{figure}

	Figure \ref{fig:Scenario} illustrates the traffic scenario \cite{Buchholz2018} we are focusing on: The ego vehicle is approaching an intersection from a minor road while two other vehicles, $\text{V}_a$ and $\text{V}_b$, approach the intersection from the main road having right of way. For the scenario, our goal is to merge either before the first car $\text{V}_a$, into the traffic gap between $\text{V}_a$ and $\text{V}_b$, or behind $\text{V}_b$. If all of these options are impossible for the given traffic situation, the ego vehicle should yield at the yield line.
	To generate safe trajectories, we settle upon Pek et al. \cite{Pek2017}, who introduced the concepts \textit{Point of No Return} (PNR) and \textit{Point of Guaranteed Arrival} (PGA): the PNR is the state $x(t_\text{PNR}) \in \mathcal{T}$ at time $t_\text{PNR}$, where returning to a state within the initial set of safe states is ultimately possible. In our case, this set of safe states only contains $x_\text{safe} = [s_\text{yield},v=0,a=0]^T$, where $s_\text{yield}$ is the position of the yield line, $v$ is the velocity and $a$ the acceleration.
	Thus, the PNRs can be directly calculated from $v(t_{PNR})$ by $x_{PNR} = [ s_{yield} - \frac{v(t_{PNR})^2}{2 b_{max}}, v(t_{PNR}), -b_{max}]^T$, where $b_{max}$ is the maximum acceptable deceleration. The fail-safe strategy then is to break with constant deceleration $b \leq b_{max}$ until the ego vehicle comes to a full stop at $s_{yield}$. Therefore, the existence of a fail-safe trajectory is guaranteed before the PNR. In contrast, for the safety-critical passageway between PNR and PGA, no fail-safe trajectory exists. Thus, the ego vehicle can only decide to pass the PNR when the remaining planning risk is sufficiently low. For the safety-critical passageway, the planning problem then is reduced to reactive planning only.
	In contrast, the PGA is a state $x(t_{\text{PGA}}) \in \mathcal{T}$ at time $t_\text{PGA}$, where the safe arrival in the goal state can be guaranteed by a suitable controller. Thus, the PGA is a suitable state to handover control to another planning module.
	For merging scenarios, a PGA has to fulfill the following properties:
	\begin{enumerate}
		\item The ego vehicle is positioned on the lane it merged in.
		\item The speed of the ego vehicle is matched to the speed of the vehicle directly ahead or to the free driving target speed, respectively. Thus, reactive maneuvers of a car following the ego vehicle can be considered traffic related rather than a direct consequence of the merging.
		\item The safety distances must be respected. A situation where the ego vehicle is in violation of the safety distances is considered \textit{dangerous}.
	\end{enumerate}
	
	Mathematically, the safety distances can be expressed as
	\begin{subequations}
		\begin{align}
		\small
		s_\text{safety, a} &= s_{\text{PGA}} + \hat{v}_{\text{\tiny V}_a} \cdot t_{\text{safety}} + s_{\text{margin}} + \frac{l_{v_a}}{2} \, ,\\
		s_\text{safety, b} &= s_{\text{PGA}} - \hat{v}_{\text{\tiny V}_b} \cdot t_{\text{safety}} + s_{\text{margin}} + \frac{l_{v_b}}{2} \, ,
		\end{align}
	\end{subequations}
	with the estimated velocities $\hat{v}_{\text{\tiny V}_a}$, $\hat{v}_{\text{\tiny V}_b}$, the safety time $t_{\text{safety}}$, the safety margin $ s_{\text{margin}}=2 \, \text{m}$, the PGA's position $s_{\text{PGA}}$, and the vehicle lengths $l_{\text{V}_a}$ and $l_{\text{V}_b}$.
	
	A commonly chosen criterion for comfort is the jerk, where a small jerk yields a high passenger comfort. Thus, the overall problem for one planning option defined by one pair of PNR and PGA can be formulated as
	\begin{subequations}
		\label{eq:problem}
		\begin{align}
		\small
		u^\ast(t) &= \arg \min_{u(t)} \{ J \} \quad \text{subject to} \\
		J &= \int_{0}^{t_f} \frac{1}{2} u(t)^2 dt + w_{\text{risk},\text{V}_a} \cdot p(\text{"} \text{V}_a \text{ dangerous"}| t_f) \nonumber \\
		&  +  w_{\text{risk}, \text{V}_b} \cdot p(\text{"} \text{V}_b \text{ dangerous"}| t_f) \, , \label{eq:OriginalCost} \\
		\dot{\boldsymbol{x}}_{\text{ego}} &= \begin{bmatrix}
		0 & 1 & 0 \\
		0 & 0 & 1 \\
		0 & 0 & 0
		\end{bmatrix} \cdot \begin{bmatrix}
		s_\text{ego} \\ v_\text{ego} \\ a_\text{ego}
		\end{bmatrix} + \begin{bmatrix} 0 \\ 0 \\ 1 \end{bmatrix} \cdot u \, , \\	
		a_\text{min} &\leq a(t) \leq a_\text{max} \, , \quad 0 \leq v(t) \leq v_\text{max} \, , \label{eq:dynamicConstraints}\\
		0 &\leq p(\text{"V}_a \text{ dangerous"}| t_f) \leq p_{\text{residual, max}} \, , \label{eq:risk1}\\	
		0 &\leq p(\text{"V}_b \text{ dangerous"}| t_f) \leq p_{\text{residual, max}} \, , \label{eq:risk2}\\
		\boldsymbol{x}(0) &= \boldsymbol{x}_{\text{ego},0} \, , \quad \boldsymbol{x}(t_f) = \begin{bmatrix} s_f & v_{f} & 0 \end{bmatrix}^T \, .
		\end{align}
	\end{subequations}
	In (\ref{eq:problem}), $u^\ast$ is the optimal input $u$, in this case the jerk, $J$ is the cost function, $t_f$ the time when the ego vehicle is supposed to reach the PGA in this planning option (i.e. the optimization horizon), $w_{\text{risk},\text{\tiny V}_a}$ and $w_{\text{risk},\text{\tiny V}_b}$ are the weights for the residual risks $p(\text{"V}_a \text{ dangerous"}| t_f)$ and $p(\text{"V}_b \text{ dangerous"}| t_f)$ of getting into a dangerous situation due to the vehicle ahead ($\text{V}_a$) or behind ($\text{V}_b$), respectively, given $t_f$. In essence, $w_{\text{risk},\text{\tiny V}_a}$ and $w_{\text{risk},\text{\tiny V}_b}$ are parameters that adjust how conservative the vehicle should decide whether to merge or not. For this work, these design parameters are chosen as $w_{\text{risk},\text{\tiny V}_a}=20$ and $w_{\text{risk},\text{\tiny V}_b}=50$.
	Furthermore, $\boldsymbol{x}_{\text{ego}}$ is the ego state consisting of the ego position $s_\text{ego}$, the ego velocity $v_\text{ego}$, and the ego acceleration $a_\text{ego}$.
	The states are constrained with minimum and maximum acceleration $a_\text{min}$, $a_\text{max}$ as well as the maximum velocity $v_\text{max}$, while the residual risks are constrained by the maximum accepted residual risk $p_{\text{residual, max}}$.
	The initial value is the current ego position $\boldsymbol{x}_{\text{ego},0}$, while the target state $\boldsymbol{x}(t_f)$ is the position $s_f$ at final velocity $v_f$ and zero acceleration. For planning options where the ego vehicle merges before the first vehicle, $s_f = s_\text{PGA}$ is the PGA position, while $v_f = v_\text{max}$ is the target velocity of the map.
	For merging into the gap between two vehicles, the target velocity $v_f = v_{\text{\tiny V}_a}$ is equal to the velocity of the vehicle ahead. Finally, the option \textit{comfortably yielding at the yield line} has the final state $\boldsymbol{x}_f = \boldsymbol{x}_\text{safe}$.
	Note that although both, comfortably yielding and the fail-safe strategy result in the same final state, the fail-safe solution space is significantly bigger, as passenger's comfort is completely disregarded.
	
	\subsection{Algorithm Overview}
	The key idea behind the presented planning scheme is to first determine possible target states that fulfill the PGA properties (see Section \ref{sec:ProblemFormulation}) with sufficiently high probability and then plan to the respective target states. From all calculated trajectories, the trajectory with minimal costs according to (\ref{eq:problem}) is chosen. If the primary target of merging cannot be reached, i.e. no valid trajectory is found, a trajectory to the yield line is planned.
	If still no valid trajectory can be found, the fail-safe strategy is applied.
	Algorithm \ref{alg:PlanningScheme} presents the big picture of the planning scheme, while the respective functions are explained in detail in the following sections.
	
	\begin{algorithm}
		\caption{Planning Scheme}
		\label{alg:PlanningScheme}
		\small
		\begin{algorithmic}[1]
			\renewcommand{\algorithmicrequire}{\textbf{Input:}}
			\renewcommand{\algorithmicensure}{\textbf{Output:}}
			\REQUIRE Object List $\mathcal{L}_o$, local digital map $\mathcal{M}_\text{local}$, \\
			$\quad$ \textit{optional:} Object predictions $\mathcal{P}_o$
			\ENSURE  Optimal Trajectory $\mathcal{T}^{\ast}$
			\IF {TrajectoryLocked == FALSE}
				\STATE $\tilde{\mathcal{L}}_o \gets$ PreprocessObjectList$(\mathcal{L}_o, \, \mathcal{M}_\text{local})$
				\IF {$\mathcal{P}_o = \emptyset$}
					\STATE $\mathcal{P}_o \gets$ PredictObjects$(\,)$
				\ENDIF
				\FORALL {$(o_i,o_{i+1}) \in \tilde{\mathcal{L}}_o,\, i \in \mathbb{N}$}
					\STATE $\mathcal{T} \gets$ EvaluateMergingOptions$(o_i, \, o_{i+1})$
				\ENDFOR
				\IF {isValid$(\mathcal{T})$== FALSE}
					\STATE  $\mathcal{T} \gets$ CalculateGentleStop$(\,)$
				\ENDIF
				\IF {isValid$(\mathcal{T})$}
					\STATE $\mathcal{T}^{\ast} \gets$ ApplyOptimalTrajectory$(\,)$
				\ELSE
					\STATE $\mathcal{T} \gets$ ApplyFailSafeStrategy$(\,)$
				\ENDIF	
				\IF {isBeyondPNR($\mathcal{T}$)}
					\STATE TrajectoryLocked = TRUE
				\ENDIF 		
			\ELSE
				\STATE $\mathcal{T}^{\ast} \gets$ ApplyLockedTrajectory$(\,)$
			\ENDIF
			\RETURN  $\mathcal{T}^{\ast}$
		\end{algorithmic}
	\end{algorithm}
	
	\subsection{Preprocessing and Prediction}
	The motion planning module takes as input the object list $\mathcal{L}_o$ from the ego vehicle's environmental model, which is generated by the prior perception module of the ego vehicle. From this object list, first of all, the relevant road users are selected by associating the road users with the lanes from the local digital map $\mathcal{M}_\text{local}$. Then, road users on lanes irrelevant to the ego vehicle's intended driving maneuver are removed from $\mathcal{L}_o$. The remaining relevant road users then are projected onto the lanes to reduce the prediction problem to 1D. Finally, the list of relevant road users is sorted with respect to their distance to the merging point. This results in the preprocessed list of road users $\tilde{\mathcal{L}}_o$.
	
	As for this paper we focus in planning, we use a prediction as simple as a Kalman filter with underlying constant velocity model and show that our planning scheme still works. The result of the prediction is represented in a very general way as tuple $(\hat{\boldsymbol{x}}, \, \begin{bmatrix} \sigma_{11} & \sigma_{12} \\ \sigma_{21} & \sigma_{22}\end{bmatrix})$, where $\hat{\boldsymbol{x}}$ is the estimated state and $ \sigma_{11}, \,  \sigma_{12}, \,  \sigma_{21}, \,  \sigma_{22} $ are its corresponding covariances. Thus, more elaborate prediction mechanisms can easily interface with our planning scheme.
	
	\subsection{Evaluation of Merging Options}
	The basic idea behind this step is to sample PGAs between the vehicles driving along the main road over time to find suitable merging options. Thus, it is iterated through $\tilde{\mathcal{L}}_o$, and two objects $o_i, \, o_{i+1}$ as well as their corresponding predictions $\mathcal{P}_{o_i}$, $\mathcal{P}_{o_{i+1}}$ are selected each iteration. Then, within $\mathcal{P}_{o_i}$, i.e. the predictions of the vehicle ahead, the smallest time $t_{f,0}$ is searched where
	\begin{equation}
	\small
	\!p(\text{"V}_a \text{ dangerous"}| t_{f,0}) = 1 - \Phi \left( \frac{\hat{s}_{o_i}(t_{f,0})-s_\text{safety, ahead}(t_{f,0}) }{\sigma_{11,o_i}^2} \right)
	\end{equation}
	fulfills the constraint (\ref{eq:risk1}). Hereby, $\Phi(\cdot)$ is the Gaussian cumulative distribution function \cite{Bronshtein2007}. From $t_{f,0}$, the target states are sampled over time until the constraint (\ref{eq:risk2}) with
	\begin{equation}
	\small
	\!p(\text{"V}_b \text{ dangerous"}| t_{f}) = 1 - \Phi \left( \frac{s_\text{safety, behind}(t_{f}) - \hat{s}_{o_{i+1}}(t_{f}) }{\sigma_{11,o_{i+1}}^2} \right)
	\end{equation}
	is not fulfilled anymore.
	
	After all iterations, this leads to the set of target point candidates
	\begin{equation}
	\small
	\mathcal{S}_{\text{target}} = \{ t_{f,j}, \, x(t_{f,j}) = [ s_{\text{PGA}}, \, \hat{v}_{o_i}(t_{f,j}), \, \\ 0	]^T \} .
	\end{equation}
	Then, a trajectory candidate is calculated connecting the current ego vehicle state $x_{\text{ego}}$ with each target state candidate $\boldsymbol{x}(t_{f,j}) \in \mathcal{S}_{\text{target}}$ using our analytical planning scheme (see Section \ref{sec:TimeWeightedJerkOptimal}). A subsequent sanity check tests whether or not the trajectory candidate fulfills the state constraints (\ref{eq:dynamicConstraints}) by sampling roughly over the trajectory. Furthermore, the last ultimate timestep is searched where the ego vehicle can stop at the yield line. The corresponding state then is the PNR of the respective trajectory. 	
	Within the evaluation of all merging options, merging before the first road user in $\tilde{\mathcal{L}}_o$, $o_1$, represents a special case with the unconstrained goal velocity $v_\text{goal} = v_\text{max}$ (see Section \ref{sec:ProblemFormulation}). Furthermore, the search direction is reverted. Thus, the time $t_{f,\text{end}}$ is searched for which the constraint (\ref{eq:risk2}) is ultimately fulfilled. From $t_{f,\text{end}}$, target state candidates are sampled over time until no valid trajectory is found anymore due to the dynamic constraints  (\ref{eq:dynamicConstraints}).
	Note that due to the optimization for the jerk, the resulting trajectory is expected to be sufficiently smooth so that constraint violations in between two sampling points can be neglected. This creates the trajectory family from which the optimal trajectory $\mathcal{T}^\ast$ can be selected by evaluating the cost (\ref{eq:OriginalCost}).
	
	\subsection{Calculating Trajectory to Gently Stop at Yield}
	If the traffic situation does not permit to merge, the evaluation of merging options will return no valid trajectory. In this case, a comfortable trajectory is planned to stop at the yield line. Hence, the target state is $\boldsymbol{x}(t_f) = [s_\text{yield}, \, 0, \, 0]^T$. As for yielding the target state is static, a solely jerk optimal trajectory is planned as described in \cite{Werling2010}. Once $\boldsymbol{x}_\text{ego} = \boldsymbol{x}(t_f)$, the merging problem is simplified to the problem formulation descirbed by Puphal et al. \cite{Puphal2018} and can be solved accordingly.
	
	\subsection{Application to the Actuators}
	After the merging options are evaluated, the optimal trajectory $\mathcal{T}^\ast$ is applied to the actuators. If, however, neither a valid trajectory for merging, nor a trajectory to comfortably stop at the yield line is found, the fail-safe strategy is applied. This means that the ego vehicle breaks with deceleration $b \leq b_\text{max}$ to get to a full stop at the yield line.
	
	As the planning scheme is based on path-velocity decomposition, the optimal trajectory $\mathcal{T}^\ast$ is passed to a subordinate control structure that can be realized, e.g. according to the approach of Graf et al. \cite{Graf2018}.
	
	\section{TRAJECTORY GENERATION} \label{sec:TrajectoryGeneration}
	In literature, the jerk is frequently used as criterion for passenger comfort, i.e. low jerk corresponds to high passenger comfort. Hence, it comes naturally to one's mind to directly calculate the jerk-optimal trajectory as proposed by Werling et al. \cite{Werling2010}. However, in a noisy environment, where the target state jumps between the time steps due to drastic changes in the predicted target states, the re-planning might still introduce a lot of jerk into the actual trajectory. To mitigate this effect, we introduce a time weight $w_t$ into the objective function
	\begin{equation}
		\small
		\label{eq:timeWeighted_J}
		\tilde{J} = \int_{0}^{t_f} \frac{1}{2} \left( \frac{w_t-1}{1+t} + 1 \right) u(t)^2 dt \, .
	\end{equation}
	Thus, control actions in the near future, which are highly likely to be actually applied to the system, are weighted stronger than control actions that are far into the future and might change due to re-planning. We demonstrate this effect in Section \ref{sec:EvalTimeWeightedJerkOptimal}, while the analytical solution to the minimization of (\ref{eq:timeWeighted_J}) is derived in Section \ref{sec:TimeWeightedJerkOptimal}.
	
	\subsection{Time Weighted Jerk Optimal Solution}\label{sec:TimeWeightedJerkOptimal}
	
	\begin{theorem}[Time-Weighted Jerk Optimal Trajectory]
		The trajectory
		\begin{align}
			\label{eq:theorem_sol}
			\small
			\mathcal{T}: \begin{bmatrix}
			\boldsymbol{x}(t) \\ u(t)
			\end{bmatrix} &= \begin{bmatrix}
				t^5   & \!\!\!\!t^4   & \!\!\!\!\!\!t^3  & \!\!\!\!-\frac{1}{4} t^2 - \frac{w_t}{2}t& \!\!\!\!t^2 & \!\!\!\!\!\!1 & \!\!\!\!\!\!0 \\
				5t^4  & \!\!\!\!4t^3  & \!\!\!\!\!\!3t^2 & \!\!\!\!-t                             & \!\!\!\!t   & \!\!\!\!\!\!1 & \!\!\!\!\!\!0 \\
				20t^3 & \!\!\!\!12t^2 & \!\!\!\!\!\!6t   & \!\!\!\!0                              & \!\!\!\!1   & \!\!\!\!\!\!0 & \!\!\!\!\!\!0 \\
				60t^2 & \!\!\!\!24t   & \!\!\!\!\!\!6    & \!\!\!\!0                              & \!\!\!\!0   & \!\!\!\!\!\!0 & \!\!\!\!\!\!0
			\end{bmatrix} \, \begin{bmatrix}\alpha_1 \\ \alpha_2 \\ \alpha_3 \\ \beta \\ \alpha_4 \\ \alpha_5 \\ \alpha_6 \end{bmatrix} \nonumber \\
			&+ \beta \, \begin{bmatrix}
			\frac{(w_t+t)^2}{2} \ln(w_t+t) \\
			(w_t+t) \ln(w_t+t) \\
			\ln(w_t+t) \\
			\frac{1}{w_t+t}
			\end{bmatrix}
		\end{align}
		solves the optimal control problem
		\begin{equation}
			\label{eq:timeWeightedOCP}
			\small
			u^\ast(t) = \arg \min_{u(t)} \{ \tilde{J} \} \; \text{subject to } \dot{\boldsymbol{x}} = \begin{bmatrix}
			0 & 1 & 0 \\
			0 & 0 & 1 \\
			0 & 0 & 0
			\end{bmatrix} \, \boldsymbol{x} + \begin{bmatrix} 0 \\ 0 \\ 1 \end{bmatrix} u \, ,
		\end{equation}
		where $\tilde{J}$ is defined according to (\ref{eq:timeWeighted_J}). As $\beta$ is a function of the coefficients $\alpha_1$, $\alpha_2$, and $\alpha_3$, the coefficients are uniquely determined through $\boldsymbol{x}(0) = \boldsymbol{x}_0$ and $\boldsymbol{x}(t_f) = \boldsymbol{x}_f$.
	\end{theorem}
	\vspace{8pt}
	\begin{proof}
		Using the Hamiltonian formulation \cite{Kamien2012}, the Hamilton function to (\ref{eq:timeWeightedOCP}) is given by
		\begin{equation}
			\small
			H(\boldsymbol{x},\boldsymbol{u},\boldsymbol{\lambda}) = \frac{1}{2} \left(\frac{w_t-1}{1+t} + 1\right)u^2 + \lambda_1 x_2 + \lambda_2 x_3 + \lambda_3 u \, .
		\end{equation}
		Thus, the conjugate states can be calculated by
		\begin{equation}
			\small
			\dot{\boldsymbol{\lambda}} = - \frac{\partial H}{\partial x} = \begin{bmatrix} 0 \\ -\lambda_2 \\ - \lambda_3 \end{bmatrix} \overset{\int \, dt}{\implies} \begin{bmatrix}
			\lambda_1 \\ \lambda_2 \\ \lambda_3
			\end{bmatrix} = \begin{bmatrix}
			c_1 \\ -c_1 t + c_2 \\ \frac{1}{2} c_1 t^2 - c_2 t + c_3
			\end{bmatrix} \, ,
		\end{equation}
		where $c_1$, $c_2$, $c_3 \in \mathbb{R}$ are integration constants. With the second Hamiltonian equation, the input is calculated:
		\begin{equation}
			\small
			\label{eq:Hamiltonian_u}
			0 = \frac{\partial H}{\partial u} \iff u = - \frac{1 + t}{w_t+t} \lambda_3 \, .
		\end{equation}
		Then, a polynomial long division is applied to (\ref{eq:Hamiltonian_u}), yielding
		\begin{align}
			\small
			u &= \underbrace{-\frac{1}{2} c_1}_{=:\tilde{\alpha}_1} t^2 + \underbrace{(c_2 - \frac{1}{2}c_1 + \frac{1}{2} c_1 w_t)}_{=: \tilde{\alpha}_2}t \nonumber \\
			  &\quad+ \underbrace{(c_2 - c_3 - w_t c_2 + \frac{1}{2} c_1 w_t - \frac{1}{2} c_1 w_t^2)}_{=:\tilde{\alpha}_3} \nonumber\\
			  &\quad+ \underbrace{\frac{w_t(c_3 - c_2 + 2 c_2 - \frac{1}{2} c_1 w_t + \frac{1}{2} c_1 w_t^2)-c_3}{w_t+t}}_{=: \frac{\beta}{w_t+t}} \, ,
		\end{align}
		where $\beta$ is a function of $c_1$, $c_2$, and $c_3$. Finally, integrating the remaining Hamiltonian differential equation
		\begin{equation}
			\small
			\dot{\boldsymbol{x}} = \frac{\partial H}{\partial \boldsymbol{\lambda}} = \begin{bmatrix}
			x_2 \\ x_3 \\ u
			\end{bmatrix} \overset{\int \, dt}{\underset{\text{subs. }\tilde{\alpha}_{1 \dots 5}}{\implies}} \text{eq. (\ref{eq:theorem_sol})}
		\end{equation}		
		with respect to $t$ and substituting $\alpha_1 = \frac{\tilde{\alpha}_1}{60}$, $\alpha_2 = \frac{\tilde{\alpha}_2}{12}$, $\alpha_3 = \frac{\tilde{\alpha}_3}{6}$, $\alpha_4 = \frac{\tilde{\alpha}_4}{2}$, $\alpha_5 = \tilde{\alpha}_5$, $\alpha_6 = \tilde{\alpha}_6$ then yields the result (\ref{eq:theorem_sol}).
	\end{proof}
	\vspace{8pt}
	\textbf{\textit{Remark:}} In order to solve (\ref{eq:theorem_sol}) such that $\boldsymbol{x}(0) = \boldsymbol{x}_0$ and $\boldsymbol{x}(t_f) = \boldsymbol{x}_f$, we first solve the set of equations in terms of $\alpha_{1 \dots 6} , \, \beta$, $\boldsymbol{x}(0) = \boldsymbol{x}_0$ and $\boldsymbol{x}(t_f) = \boldsymbol{x}_f$ which yields an under-determined yet linear set of equations.
	We solve this linear set of equations for $\alpha_{1 \dots 6}$ in dependence of $\beta$ as free variable. Then, we express the integration constants $c_1$, $c_2$, and $c_3$ in terms of $\alpha_1$, $\alpha_2$ and $\alpha_3$. Thus, $\beta$ can be determined through the relation
	\begin{equation}
	\small
	\beta = w_t(c_3 - c_2 + 2 c_2 - \frac{1}{2} c_1 w_t + \frac{1}{2} c_1 w_t^2)-c_3 \, ,
	\end{equation}
	$\alpha_1$, $\alpha_2$ and $\alpha_3$ in turn can be expressed as functions of $\boldsymbol{x}_0$, $\boldsymbol{x}_f$ and $\beta$, thus $\beta$ can be determined depending on $\boldsymbol{x}_0$ and $\boldsymbol{x}_f$ only.
	We omit them here, while these can be calculated easily by use of a computer algebra system, e.g. MATLAB Symbolic Toolbox \cite{MATLAB2017}.

	\subsection{Comparison with Jerk Optimal Solution} \label{sec:EvalTimeWeightedJerkOptimal}
\begin{figure}[thpb]
		\centering
		\begin{tikzpicture}
		\footnotesize
		\input{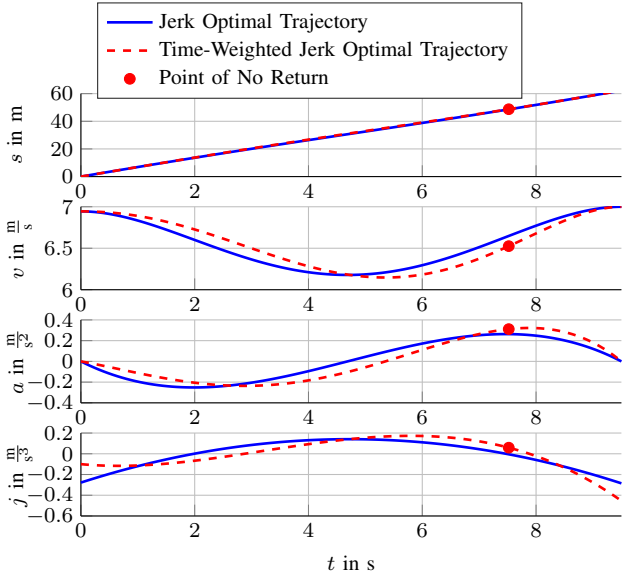}
		\end{tikzpicture}
		\caption{Jerk optimal and time-weighted jerk optimal trajectory in comparison for a single shot. The trajectory consists of the states position $s$, velocity $v$, acceleration $a$ and jerk $j$ over time $t$.}
		\label{fig:Trajectories}
	\end{figure}

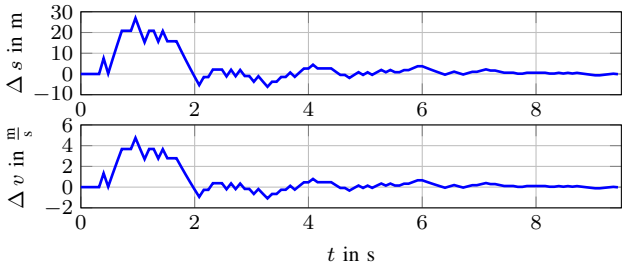
\begin{figure}[thpb]
		\centering
		\begin{tikzpicture}
		\footnotesize
%\begin{tikzpicture}

\begin{axis}[%
width=0.4\textwidth,
height=1.1cm,
at={(0.0cm,1.5cm)},
scale only axis,
xmin=0,
xmax=9.5,
ymin=-10,
ymax=30,
xmajorgrids,
ymajorgrids,
y label style={at={(axis description cs:0.05,0.5)}},
ylabel={$\Delta \, s \text{ in m}$},
axis background/.style={fill=white}
]
\addplot [color=blue, line width=1.0pt, forget plot]
  table[row sep=crcr]{%
0	0\\
0.08	0\\
0.16	0\\
0.24	0\\
0.32	0\\
0.4	7.524988885\\
0.48	0.225749666550001\\
0.56	7.3060117084465\\
0.64	14.1738658890861\\
0.72	20.8356844443065\\
0.8	20.8356844443065\\
0.88	20.8356844443065\\
0.96	26.9157463705552\\
1.04	21.018086302094\\
1.12	15.2973560356866\\
1.2	20.8464643941018\\
1.28	20.8464643941018\\
1.36	15.6253083396689\\
1.44	20.6898297124688\\
1.52	15.7772439808529\\
1.6	15.7772439808529\\
1.68	15.7772439808529\\
1.76	11.2936596234219\\
1.84	6.9445827967138\\
1.92	2.72597827480695\\
2	-1.3660681114427\\
2.08	-5.33535310610485\\
2.16	-1.48514666128256\\
2.24	-1.48514666128256\\
2.32	2.13751258265073\\
2.4	2.13751258265073\\
2.48	2.13751258265073\\
2.56	-1.1687906974876\\
2.64	2.03832348424658\\
2.72	-1.07257727203557\\
2.8	1.94499646155812\\
2.88	-0.982050060027762\\
2.96	-0.982050060027762\\
3.04	-3.73610813218792\\
3.12	-1.06467180219257\\
3.2	-3.65596504228806\\
3.28	-6.16951948518068\\
3.36	-3.73137167557483\\
3.44	-3.73137167557483\\
3.52	-1.43731840151669\\
3.6	-1.43731840151669\\
3.68	0.72115632404461\\
3.76	-1.37256415974985\\
3.84	0.658344709530776\\
3.92	2.62832631273299\\
4	2.62832631273299\\
4.08	4.48188200318595\\
4.16	2.68393298344658\\
4.24	2.68393298344658\\
4.32	2.68393298344658\\
4.4	2.68393298344658\\
4.48	1.09222164352476\\
4.56	-0.451738356199397\\
4.64	-0.451738356199397\\
4.72	-1.90445031993986\\
4.8	-0.495319715111611\\
4.88	0.87153697157179\\
4.96	-0.454314014511109\\
5.04	0.831761441989303\\
5.12	-0.415731750816096\\
5.2	0.794336646205141\\
5.28	1.96810299131574\\
5.36	0.829549636558459\\
5.44	1.93394639067302\\
5.52	0.862681539181896\\
5.6	0.862681539181896\\
5.68	1.8706346379499\\
5.76	1.8706346379499\\
5.84	2.81901770858071\\
5.92	3.7389492870926\\
6	3.7389492870926\\
6.08	2.87338566487076\\
6.16	2.03378895131558\\
6.24	1.21938013916706\\
6.32	0.429403591382989\\
6.4	-0.336873659967558\\
6.48	0.406415273842472\\
6.56	1.1274055396382\\
6.64	0.428044981816344\\
6.72	-0.250334759270858\\
6.8	0.407693589583728\\
6.88	1.04598108797268\\
6.96	1.04598108797268\\
7.04	1.64654579520684\\
7.12	2.22909356122397\\
7.2	1.66402222818735\\
7.28	1.66402222818735\\
7.36	1.13234661093319\\
7.44	0.61662126219666\\
7.52	0.61662126219666\\
7.6	0.61662126219666\\
7.68	0.145932660989243\\
7.76	0.145932660989243\\
7.84	0.588803565865302\\
7.92	0.588803565865302\\
8	0.588803565865302\\
8.08	0.588803565865302\\
8.16	0.588803565865302\\
8.24	0.208495250855682\\
8.32	0.208495250855682\\
8.4	0.566327344448233\\
8.48	0.219230213663458\\
8.56	0.55591443052469\\
8.64	0.229330740169295\\
8.72	0.546116919814028\\
8.8	0.238834325558637\\
8.88	-0.0592297908690921\\
8.96	-0.348351983803989\\
9.04	-0.62880051095084\\
9.12	-0.62880051095084\\
9.2	-0.364926491758368\\
9.28	-0.108968693141671\\
9.36	0.139310371516525\\
9.44	-0.101520321201925\\
};
\end{axis}

\begin{axis}[%
width=0.4\textwidth,
height=1.1cm,
at={(0.0cm,0.0cm)},
scale only axis,
xmin=0,
xmax=9.5,
x label style={at={(axis description cs:0.5,0.1)}},
xlabel={$t$ in s},
ymin=-2,
ymax=6,
xmajorgrids,
ymajorgrids,
y label style={at={(axis description cs:0.06,0.5)}},
ylabel={$\Delta \, v \text{ in } \frac{\text{m}}{\text{s}}$},
axis background/.style={fill=white}
]
\addplot [color=blue, line width=1.0pt, forget plot]
  table[row sep=crcr]{%
0	0\\
0.08	0\\
0.16	0\\
0.24	0\\
0.32	0\\
0.4	1.327939215\\
0.48	0.03983817645\\
0.56	1.2892961838435\\
0.64	2.50127045101519\\
0.72	3.67688549017174\\
0.8	3.67688549017174\\
0.88	3.67688549017174\\
0.96	4.74983759480386\\
1.04	3.7090740533107\\
1.12	2.69953341806234\\
1.2	3.67878783425325\\
1.28	3.67878783425325\\
1.36	2.75740735405922\\
1.44	3.65114641984743\\
1.52	2.78421952603287\\
1.6	2.78421952603287\\
1.68	2.78421952603287\\
1.76	1.99299875707445\\
1.84	1.22551461118479\\
1.92	0.481054989671814\\
2	-0.241070843195771\\
2.08	-0.941532901077329\\
2.16	-0.262084704932218\\
2.24	-0.262084704932218\\
2.32	0.377208102820717\\
2.4	0.377208102820717\\
2.48	0.377208102820717\\
2.56	-0.206257181909577\\
2.64	0.359704144278808\\
2.72	-0.189278342123926\\
2.8	0.343234669686726\\
2.88	-0.173302951769606\\
2.96	-0.173302951769606\\
3.04	-0.659313199797869\\
3.12	-0.187883259210454\\
3.2	-0.645170301580247\\
3.28	-1.08873873267895\\
3.36	-0.658477354513208\\
3.44	-0.658477354513208\\
3.52	-0.253644423797065\\
3.6	-0.253644423797065\\
3.68	0.127262880713754\\
3.76	-0.24221720466174\\
3.84	0.116178478152489\\
3.92	0.463822290482291\\
4	0.463822290482291\\
4.08	0.790920353503402\\
4.16	0.473635232372925\\
4.24	0.473635232372925\\
4.32	0.473635232372925\\
4.4	0.473635232372925\\
4.48	0.192744995916134\\
4.56	-0.0797185334469535\\
4.64	-0.0797185334469535\\
4.72	-0.336079468224682\\
4.8	-0.0874093614902853\\
4.88	0.15380064204208\\
4.96	-0.0801730613843144\\
5.04	0.146781430939288\\
5.12	-0.0733644266146063\\
5.2	0.140177055212671\\
5.28	0.34731229258513\\
5.36	0.146391112333845\\
5.44	0.341284657177591\\
5.52	0.152237918679157\\
5.6	0.152237918679157\\
5.68	0.330111994932334\\
5.76	0.330111994932334\\
5.84	0.497473713278948\\
5.92	0.659814580075163\\
6	0.659814580075163\\
6.08	0.507068058506604\\
6.16	0.358903932585102\\
6.24	0.215184730441244\\
6.32	0.0757771043617028\\
6.4	-0.0594482929354525\\
6.48	0.0717203424427882\\
6.56	0.198953918759682\\
6.64	0.075537349732295\\
6.72	-0.0441767222242701\\
6.8	0.071945927573598\\
6.88	0.18458489787753\\
6.96	0.18458489787753\\
7.04	0.2905669050365\\
7.12	0.3933694519807\\
7.2	0.293650981444826\\
7.28	0.293650981444826\\
7.36	0.199825872517621\\
7.44	0.108815516858233\\
7.52	0.108815516858233\\
7.6	0.108815516858233\\
7.68	0.0257528225275124\\
7.76	0.0257528225275124\\
7.84	0.103906511623288\\
7.92	0.103906511623288\\
8	0.103906511623288\\
8.08	0.103906511623288\\
8.16	0.103906511623288\\
8.24	0.0367932795627664\\
8.32	0.0367932795627664\\
8.4	0.0999401196085108\\
8.48	0.0386876847641387\\
8.56	0.0981025465631796\\
8.64	0.0404701306181099\\
8.72	0.0963735740848275\\
8.8	0.0421472339221115\\
8.88	-0.0104523160357231\\
8.96	-0.0614738794948227\\
9.04	-0.110964796050149\\
9.12	-0.110964796050149\\
9.2	-0.0643987926632425\\
9.28	-0.0192297693779429\\
9.36	0.0245841832087976\\
9.44	-0.0179153508003407\\
};
\end{axis}
%\end{tikzpicture}%
		\end{tikzpicture}
		\caption{Noisy target state. Due to prediction uncertainty, the initial target state is changed by the position offset $\Delta s$ and the velocity offset $\Delta v$, while the target acceleration is always $a=0 \frac{\text{m}}{\text{s}^2}$}.
		\label{fig:TargetState}
	\end{figure}

\begin{figure}[thpb]
		\centering
		\begin{tikzpicture}
		%\normalsize
		\footnotesize
		\input{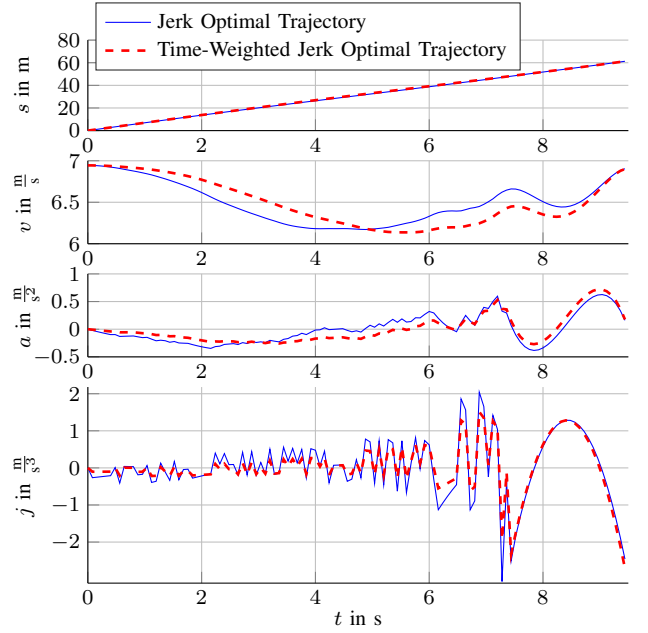}
		\end{tikzpicture}
		\caption{Comparison of actual trajectories after re-planning. The trajectory consists of the states position $s$, velocity $v$, acceleration $a$, and jerk $j$ over time $t$.}
		\label{fig:TrajectoryComparison}
	\end{figure}

	In this section, the time-weighted jerk optimal trajectories are compared with the jerk optimal trajectories corresponding to $w_t=1$. Figure \ref{fig:Trajectories} shows the time-weighted trajectory in direct comparison with the jerk optimal trajectory for a single shot planning. It can be seen that the jerk for the time-weighted trajectory is smaller at first, but higher close to $t_f$. This is the expected behavior. In the next step, noise is added to the target state (see Fig. \ref{fig:TargetState}) and the actual trajectories are compared (see Fig. \ref{fig:TrajectoryComparison}). It shows that the bumps in the jerk resulting from the replanning are smaller in the actual trajectory with time weighting as compared to the trajectory without time weighting. Note that for $t > 7.44\, \text{s}$, the target state is locked and the remaining trajectory is planned solely jerk-optimal. This corresponds to the situation in the PNR, where time weighting is not useful any longer, as the trajectory is locked and no re-planning takes place until the PGA is reached.
	
	\section{SIMULATIONS}
	In order to show the capabilities of our proposed planning scheme, we evaluate it through Monte-Carlo simulation. Our goal is to show the decision making capabilities of our approach and its ability to efficiently merge into sufficiently wide traffic gaps in various traffic situation for the scenario of an unsignalized intersection as described in Section \ref{sec:ProblemFormulation}. Furthermore, we want to demonstrate its real-time capability. The traffic situation is based on the pilot side described in \cite{Buchholz2018} and the parameters of the simulation are based on observations from the real traffic at that intersection.
	
\begin{figure}[thpb]
		\centering
		\begin{tikzpicture}
		%\normalsize
		\footnotesize
		\input{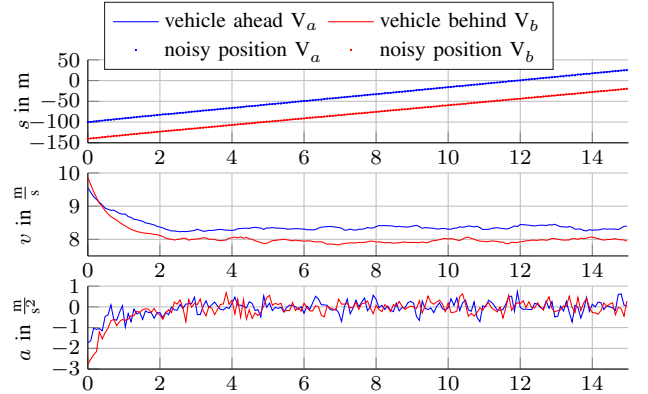}
		\end{tikzpicture}
		\caption{Trajectories of the other vehicles.}
		\label{fig:ExampleOtherVehicles}
	\end{figure}

	Through the Monte-Carlo simulation, the trajectories of the two oncoming vehicles are modeled with the intelligent driver model (IDM) \cite{Treiber2000}. In order to account for variations in the driving behavior, Gaussian noise with a standard deviation of $\sigma_a = 0.25 \, \frac{\text{m}}{\text{s}^2}$ is added to the IDM acceleration in each sampling point individually. The other vehicles' dynamics are modeled in 1D as a double integrator assuming that they follow their lane.
	Furthermore, Gaussian noise with a standard deviation of $\sigma_s = 0.25 \, \text{m}$ is added to the vehicles' positions to account for measurement uncertainties in the ego vehicle's perception. Furthermore, the arrival time of the other vehicles at the merging point is randomly chosen within the interval $[5, \, 13] \, \text{s}$.
	The initial velocities of the other vehicles are chosen as $\frac{30}{3.6} \, \frac{\text{m}}{\text{s}}$, then a Gaussian noise with the standard deviation $\sigma_v = 0.3 \frac{\text{m}}{\text{s}}$ is added.
	Figure \ref{fig:ExampleOtherVehicles} exemplarily shows the  other vehicles' trajectories for one Monte-Carlo simulation run.
	
	For the ego vehicle, the starting position is always the same and chosen such that mostly only the two merging options \textit{merging before first vehicle} and \textit{merging into gap between the vehicles} are possible. If the ego vehicle decides to merge behind the second car, this is counted as \textit{merging into gap between the vehicles}, as the last vehicle spans a gap between its back and the end of sight. If merging is impossible, the ego vehicle has to yield at the yield line. The initial velocity of the ego vehicle is varied randomly within the interval $[\frac{25}{3.6}, \, \frac{35}{3.6}] \, \frac{\text{m}}{\text{s}}$.
	
	The initial traffic gap size is swept from $30 \, \text{m}$ to $65 \, \text{m}$ in steps of $5 \, \text{m}$, while $1000$ simulation runs are conducted for each setting.
	
	\subsection{Evaluation of Planning Decisions}
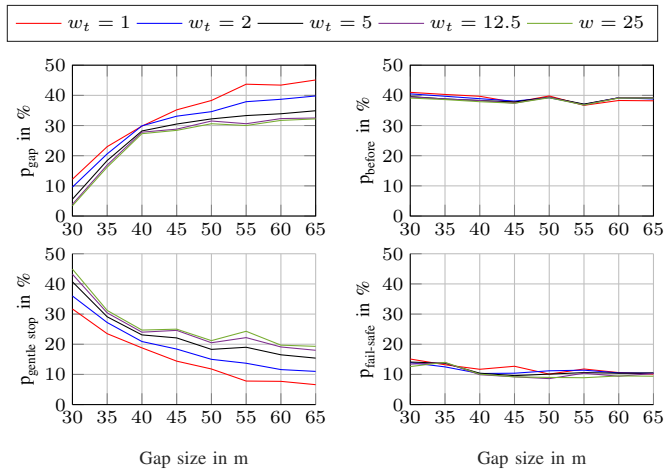
\begin{figure}[thpb]
		\centering
		\begin{tikzpicture}
		\scriptsize
		\definecolor{mycolor1}{rgb}{0.49400,0.18400,0.55600}%
		\definecolor{mycolor2}{rgb}{0.46600,0.67400,0.18800}%		
%\begin{tikzpicture}

\begin{axis}[%
width = 0.18 \textwidth,
height= 2.0cm,
at={(0.0cm,2.5cm)},
scale only axis,
xmin=30,
xmax=65,
xtick={30, 35, 40, 45, 50, 55, 60, 65},
ytick={0, 10, 20, 30, 40, 50},
xlabel style={font=\color{white!15!black}},
xlabel={Gap size in m},
ymin=0,
ymax=50,
y label style={at={(axis description cs:0.20,0.5)}},
ylabel={$\text{p}_{\text{gap}}\text{ in \%}$},
axis background/.style={fill=white},
xmajorgrids,
ymajorgrids,
legend columns=5,
legend style={at={(-0.27,1.4)}, anchor=north west, legend cell align=left, align=left, draw=white!15!black}
]
\addplot [color=red]
  table[row sep=crcr]{%
30	12.2\\
35	23\\
40	29.8\\
45	35.2\\
50	38.3\\
55	43.7\\
60	43.4\\
65	45.1\\
};
\addlegendentry{$w_t=1$}

\addplot [color=blue]
  table[row sep=crcr]{%
30	9.6\\
35	20.6\\
40	29.9\\
45	33.1\\
50	34.6\\
55	37.9\\
60	38.7\\
65	39.8\\
};
\addlegendentry{$w_t=2$}

\addplot [color=black]
  table[row sep=crcr]{%
30	5.6\\
35	18.4\\
40	28.2\\
45	30.5\\
50	32.2\\
55	33.3\\
60	33.9\\
65	34.9\\
};
\addlegendentry{$w_t=5$}

\addplot [color=mycolor1]
  table[row sep=crcr]{%
30	3.9\\
35	16.9\\
40	27.8\\
45	28.8\\
50	31.5\\
55	30.6\\
60	32.3\\
65	32.5\\
};
\addlegendentry{$w_t=12.5$}

\addplot [color=mycolor2]
  table[row sep=crcr]{%
30	3.4\\
35	16.2\\
40	27.3\\
45	28.4\\
50	30.6\\
55	30\\
60	31.7\\
65	32.2\\
};
\addlegendentry{$w=25$}

\end{axis}

\begin{axis}[%
width=0.18\textwidth,
height=2.0cm,
at={(0.25\textwidth,2.5cm)},
scale only axis,
xmin=30,
xmax=65,
xtick={30, 35, 40, 45, 50, 55, 60, 65},
ytick={0, 10, 20, 30, 40, 50},
xlabel style={font=\color{white!15!black}},
xlabel={Gap size in m},
ymin=0,
ymax=50,
y label style={at={(axis description cs:0.20,0.5)}},
ylabel={$\text{p}_{\text{before}}\text{ in \%}$},
axis background/.style={fill=white},
xmajorgrids,
ymajorgrids
]
\addplot [color=red, forget plot]
  table[row sep=crcr]{%
30	41\\
35	40.3\\
40	39.7\\
45	37.7\\
50	39.8\\
55	36.7\\
60	38.3\\
65	38.2\\
};
\addplot [color=blue, forget plot]
  table[row sep=crcr]{%
30	40.5\\
35	39.7\\
40	38.9\\
45	38.1\\
50	39.2\\
55	37\\
60	39.2\\
65	38.8\\
};
\addplot [color=black, forget plot]
  table[row sep=crcr]{%
30	39.6\\
35	38.8\\
40	38.3\\
45	37.8\\
50	39.4\\
55	37.1\\
60	39.2\\
65	39.2\\
};
\addplot [color=mycolor1, forget plot]
  table[row sep=crcr]{%
30	39.5\\
35	38.8\\
40	38.3\\
45	37.4\\
50	39.4\\
55	36.8\\
60	39.2\\
65	39.2\\
};
\addplot [color=mycolor2, forget plot]
  table[row sep=crcr]{%
30	39.1\\
35	38.6\\
40	37.9\\
45	37.4\\
50	39.2\\
55	36.8\\
60	39.1\\
65	39.1\\
};
\end{axis}

\begin{axis}[%
width = 0.18 \textwidth,
height= 2.0cm,
at={(0.0cm,0.0cm)},
scale only axis,
xmin=30,
xmax=65,
xtick={30, 35, 40, 45, 50, 55, 60, 65},
ytick={0, 10, 20, 30, 40, 50},
xlabel style={font=\color{white!15!black}},
xlabel={Gap size in m},
ymin=0,
ymax=50,
y label style={at={(axis description cs:0.20,0.5)}},
ylabel={$\text{p}_{\text{gentle stop}}\text{ in \%}$},
axis background/.style={fill=white},
axis x line*=bottom,
axis y line*=left,
xmajorgrids,
ymajorgrids
]
\addplot [color=red, forget plot]
  table[row sep=crcr]{%
30	31.7\\
35	23.5\\
40	18.8\\
45	14.4\\
50	11.8\\
55	7.8\\
60	7.7\\
65	6.6\\
};
\addplot [color=blue, forget plot]
  table[row sep=crcr]{%
30	36\\
35	27.2\\
40	20.9\\
45	18.4\\
50	15\\
55	13.7\\
60	11.6\\
65	11\\
};
\addplot [color=black, forget plot]
  table[row sep=crcr]{%
30	40.7\\
35	29.1\\
40	23.1\\
45	22.1\\
50	18.3\\
55	19\\
60	16.5\\
65	15.4\\
};
\addplot [color=mycolor1, forget plot]
  table[row sep=crcr]{%
30	43.2\\
35	30.4\\
40	24\\
45	24.6\\
50	20.5\\
55	22.2\\
60	19.1\\
65	18\\
};
\addplot [color=mycolor2, forget plot]
  table[row sep=crcr]{%
30	44.9\\
35	31.2\\
40	24.7\\
45	25\\
50	21.2\\
55	24.3\\
60	19.7\\
65	19.3\\
};
\end{axis}

\begin{axis}[%
width = 0.18 \textwidth,
height= 2.0cm,
at={(0.25 \textwidth,0.0cm)},
scale only axis,
xmin=30,
xmax=65,
xtick={30, 35, 40, 45, 50, 55, 60, 65},
ytick={0, 10, 20, 30, 40, 50},
xlabel style={font=\color{white!15!black}},
xlabel={Gap size in m},
ymin=0,
ymax=50,
y label style={at={(axis description cs:0.20,0.5)}},
ylabel={$\text{p}_{\text{fail-safe}}\text{ in \%}$},
axis background/.style={fill=white},
axis x line*=bottom,
axis y line*=left,
xmajorgrids,
ymajorgrids
]
\addplot [color=red, forget plot]
  table[row sep=crcr]{%
30	15.1\\
35	13.2\\
40	11.7\\
45	12.7\\
50	10.1\\
55	11.8\\
60	10.6\\
65	10.1\\
};
\addplot [color=blue, forget plot]
  table[row sep=crcr]{%
30	13.9\\
35	12.5\\
40	10.3\\
45	10.4\\
50	11.2\\
55	11.4\\
60	10.5\\
65	10.4\\
};
\addplot [color=black, forget plot]
  table[row sep=crcr]{%
30	14.1\\
35	13.7\\
40	10.4\\
45	9.6\\
50	10.1\\
55	10.6\\
60	10.4\\
65	10.5\\
};
\addplot [color=mycolor1, forget plot]
  table[row sep=crcr]{%
30	13.4\\
35	13.9\\
40	9.9\\
45	9.2\\
50	8.6\\
55	10.4\\
60	9.4\\
65	10.3\\
};
\addplot [color=mycolor2, forget plot]
  table[row sep=crcr]{%
30	12.6\\
35	14\\
40	10.1\\
45	9.2\\
50	9\\
55	8.9\\
60	9.5\\
65	9.4\\
};
\end{axis}
%\end{tikzpicture}%
		\end{tikzpicture}
		\caption{Statistics on merging decision in dependence of the traffic gap between $v_a$ and $v_b$ and the time weight $w$.}
		\label{fig:MergingDecisions}
	\end{figure}

	Figure \ref{fig:MergingDecisions} shows the statistics on the merging decisions as function of the traffic gap size for different parameters of $w_t$. In order to keep comparability, the trajectories of the other vehicles for different choices of $w_t$ are kept the same. It shows that for small values of $w_t$, the planning is less defensive, hence, merging is possible more often, but the fail-safe strategy has to be applied more often as well. Small values of $w_t$ are preferable to achieve a high probability $p_\text{gap}$ to merge into a traffic gap, while, as shown in Section \ref{sec:EvalTimeWeightedJerkOptimal}, small values of $w_t$ result in more bumpy trajectories.
	
	The probability of merging before the first car $p_\text{before}$ is nearly constant, thus it shows that $p_\text{before}$ is independent of $w_t$ and the gap size.
	Furthermore, it shows that $p_\text{fail-safe}$ stabilizes at about $10\%$. This can be explained as there is an area in the state space, where the ego vehicle does not find a jerk optimal trajectory to the yield line anymore, while it is still far away from the PNR. If in such a situation $\text{V}_b$ accelerates so that the traffic gap between $\text{V}_a$ and $\text{V}_b$ is predicted to close, the vehicle has to apply the fail-safe strategy. Note, however, that the average applied fail-safe deceleration was $b_\text{avg, fail-safe} = 3.14 \frac{\text{m}}{\text{s}^2}$, while the maximum applied deceleration was $b_\text{avg, fail-safe} = 3.7 \frac{\text{m}}{\text{s}^2}$.
	For validation, all simulated trajectories, in total several 10000, have been checked for collisions. According to the test, none of the simulated trajectories showed any collision.
	
	\subsection{Evaluation of Real-Time Capability}
	Averaging over all Monte-Carlo simulations, on an Intel Xeon E5-1630v4 processor our planning scheme yields an average calculation time of $2.8 \, \text{ms}$ for merging before $\text{V}_a$, $3.5 \, \text{ms}$ for merging into the gap, and $16.0 \, \text{ms}$ for gently stopping at the yield line. Gently stopping has the highest calculation time, as the stopping trajectory is calculated after the merging options have been explored. All calculation times are far below $100 \, \text{ms}$, hence the planning scheme can be considered real-time capable.
	
	\section{CONCLUSIONS}
	In this work, a new planning scheme for merging scenarios was presented that considers multiple merging options, minimizes the risk to end up in a dangerous situation and optimizes for the passenger's comfort. To mitigate the effect of uncertain target states, an analytical solution for generating time-weighted jerk optimal trajectories was derived and its benefits were demonstrated in simulation. Furthermore, simulations showed that the planning scheme is real-time capable and able to find reasonable planning decisions.
	
	In future work, we want to implement the planning scheme on our automated vehicle to validate our results through experiment.
	
	\addtolength{\textheight}{-12cm}   % This command serves to balance the column lengths
	
	%%%%%%%%%%%%%%%%%%%%%%%%%%%%%%%%%%%%%%%%%%%%%%%%%%%%%%%%%%%%%%%%%%%%%%%%%%%%%%%%

	%%%%%%%%%%%%%%%%%%%%%%%%%%%%%%%%%%%%%%%%%%%%%%%%%%%%%%%%%%%%%%%%%%%%%%%%%%%%%%%%

	%%%%%%%%%%%%%%%%%%%%%%%%%%%%%%%%%%%%%%%%%%%%%%%%%%%%%%%%%%%%%%%%%%%%%%%%%%%%%%%%
	%\section*{APPENDIX}
	%\section*{ACKNOWLEDGMENT}
	%
	%The preferred spelling of the word ÒacknowledgmentÓ in America is without an ÒeÓ after the ÒgÓ. Avoid the stilted expression, ÒOne of us (R. B. G.) thanks . . .Ó  Instead, try ÒR. B. G. thanksÓ. Put sponsor acknowledgments in the unnumbered footnote on the first page.

	%%%%%%%%%%%%%%%%%%%%%%%%%%%%%%%%%%%%%%%%%%%%%%%%%%%%%%%%%%%%%%%%%%%%%%%%%%%%%%%%
	
	\bibliographystyle{IEEEtran}
	{
		\bibliography{ITSC2019B}}

\end{document}